\documentclass[lettersize,journal]{IEEEtran}
\usepackage{amsmath,amsfonts}
\usepackage{algorithmic}
\usepackage{algorithm}
\usepackage{array}
\usepackage[caption=false,font=normalsize,labelfont=sf,textfont=sf]{subfig}
\usepackage{textcomp}
\usepackage{stfloats}
\usepackage{url}
\usepackage{verbatim}
\usepackage{graphicx}
\usepackage{cite}
\usepackage{dsfont}
\usepackage{xcolor}
\usepackage{subfig}
\allowdisplaybreaks
\begin{document}    
\setlength{\abovedisplayskip}{3pt}
\setlength{\belowdisplayskip}{3pt}
\title{A Practical and Simple Detection and
Identification Scheme for RIS-Assisted
Systems\\


}
\author{Aymen Khaleel,~\IEEEmembership{Member,~IEEE}, Recep Vural, Mehmet C. Ilter,~\IEEEmembership{Senior Member,~IEEE}, Majid Gerami, and Ertugrul Basar,~\IEEEmembership{Fellow,~IEEE}
\thanks{During this work, Aymen Khaleel, Recep Vural, and Ertugrul Basar were with the Communications Research and Innovation Laboratory
(CoreLab), Department of Electrical and Electronics Engineering, Ko\c{c}
University, 34450 Sariyer, Istanbul. Mehmet C. Ilter was with Huawei Lund Research Center, 223 69 Lund, Sweden.

Currently, Aymen Khaleel is with the Faculty of Electrical Engineering and Information Technology, Ruhr-University Bochum, Bochum 44801, Germany. (e-mail: aymen.khaleel@rub.de). Recep Vural, Mehmet C. Ilter, and Ertugrul Basar are with the Department of Electrical Engineering, Tampere University, 33720 Tampere, Finland (email: recep.vural@tuni.fi, mehmet.ilter@tuni.fi, ertugrul.basar@tuni.fi).





Majid Gerami is with Huawei Lund Research Center, 223 69 Lund, Sweden. (email: majid.gerami@huawei.com). }
}


\maketitle
\begin{abstract}
Reconfigurable intelligent surface (RIS)-empowered communication is one of the promising physical layer enabling technologies for the upcoming sixth generation (6G) wireless networks due to their unprecedented capabilities in shaping the wireless communication environment. In this paper, we consider the unique problem of RIS identification in a mobile wireless network where multiple RISs are deployed to assist the base station (BS)-user equipment (UE) communication. Here, considering dynamic link blockages, we aim to enable the BS to perceive the UE-RIS potential associations for better resource allocation. Specifically, we first introduce a novel network-level problem where the BS aims to detect and uniquely identify RISs that are reachable by a specific UE (UE-RIS-BS link is available) in a given time slot. Next, to solve this problem, we propose a novel RIS identification and detection (RIS-ID) scheme that enables the BS to pair UEs with their corresponding reachable RISs in a given time slot. On the BS side, the proposed RIS-ID scheme can be used as an initial and essential step before optimizing each RIS to serve its nearby UE, for more efficient resource allocation. Furthermore, to assess the proposed RIS-ID scheme, we propose two performance metrics: the false and miss-detection probabilities. These probabilities are analytically derived and verified through computer simulations, revealing the effectiveness of the proposed RIS-ID scheme under different operating scenarios.
\end{abstract}

\begin{IEEEkeywords}
Reconfigurable intelligent surfaces, detection, identification.
\end{IEEEkeywords}
\section{INTRODUCTION}
\IEEEPARstart{R}{econfigurable} intelligent surfaces (RISs) continue receiving growing attention as one of the game-changing physical layer enabling technologies in the upcoming sixth generation (6G) of wireless communications. RISs composed of dynamically controllable metamaterial elements, carries the potential to bring huge enhancements into wireless communication systems through its remarkable ability to exert unprecedented control over electromagnetic waves \cite{WirelessComm_Through_RIS}. By manipulating the phase, amplitude, and polarization of signals, a RIS can enhance communication links, improve energy efficiency, and mitigate interference \cite{RIS_for_6G}. RISs have been studied and applied with various wireless communication contexts, including non-orthogonal multiple access (NOMA) \cite{RIS-NOMA}, multiple-input multiple-output (MIMO) \cite{RIS-MIMO}, index modulation \cite{RIS-IM}, and beyond. Due to their relatively low cost compared, for example, to BS-based micro cells \cite{low-cost}, RISs are envisioned to be deployed in large numbers as a cost-efficient solution for wireless coverage enhancement \cite{ris-deploy}.

Additionally, RIS technology extends beyond information communication systems and finds utility in localization \cite{wymeersch2020radio} and sensing. For instance, in \cite{hu2020RIS_Sensing}, the authors designed a radio-frequency (RF) sensing system leveraging RISs for human posture recognition where optimal RIS configuration is obtained to minimize posture recognition errors.
In \cite{Zhang_Hybrid_RIS}, the authors introduced hybrid RISs (HRISs), which not only reflect impinging signals but also incorporate active reception elements for sensing and processing. The findings indicate that HRISs enable more efficient channel estimation, requiring notably fewer pilots than fully reflective RISs.
In \cite{Liu_ISAC}, the authors investigated RIS-aided integrated sensing and communication (ISAC) within a multi-user multiple-input single-output (MISO) downlink communication system. The authors developed an algorithm for joint transmit beamforming and RIS configuration to maximize the sum-rate of multiple users while ensuring the minimum signal-to-noise ratio (SNR) required for effective target detection in the radar.
The study of  \cite{NOMA-RIS-ISAC} incorporated NOMA technology into an RIS-assisted ISAC system, where the RIS was utilized to create a virtual line-of-sight (LoS) connection between the base station (BS) and both the communication users and radar targets.

In \cite{STAR-RIS-ISAC}, the authors investigated a simultaneous transmitting and reflecting (STAR) RIS-aided ISAC system, that removes the necessity of communication users or radar targets to be positioned on the same side as the BS and RIS. In \cite{modulated}, a localization approach is introduced employing modulated RISs. Here, a BS transmits an unmodulated carrier signal, subsequently modulated by multiple RISs by applying phase shifts following an m-sequence. Signals from different RISs are distinguished thanks to their unique m-sequences and the use of a time division method to run a single RIS at a time. Note that an m-sequence is a pseudorandom binary sequence that is widely used in signal detection due to its pseudo-randomness and correlation properties. Finally, the user estimates its location using time-difference-of-arrival information from received signals. In \cite{Self_loc}, a study on BS-free RIS-aided localization is conducted. In the considered scenario, the UE estimates its location by transmitting orthogonal frequency-division multiplexing (OFDM) pilots and then analyzing the signals reflected from the RIS, whose location, position, and phase profile are known by the UE. 

Considering the application of detection and identification concepts to RIS-assisted systems, only a few related works are found in the literature. Specifically, considering an ISAC system, the authors in \cite{RIS-ID-Rel3} proposed an alternation scheme for the RIS response to enable the UE to identify RIS-assisted paths and distinguish them from non-LoS paths. In \cite{RIS-ID-Rel2}, the authors investigated the problem of detecting the presence of non-cooperative behavior of RISs in a MIMO system. 

\begin{table*}[t!]
\caption{Relevance to Existing Works}
\label{table_example}
\centering
\begin{tabular}{|c|c|c|c|c|}
\hline
\textbf{Ref.} & \textbf{Considers RIS-ID} & \textbf{Based on This Work} & \textbf{Published After This Work}\\
\hline
\cite{RIS-ID-Rel3} &  $\times$ &  $\times$&$\checkmark$ \\
\hline
\cite{RIS-ID-Rel2} & $\times$ &  $\times$ &  $\times$ \\
\hline
\cite{RIS-ID-pract} & $\checkmark$ & $\checkmark$ & $\checkmark$\\
\hline
\cite{RIS-ID-beam} & $\checkmark$ & $\checkmark$ & $\checkmark$ \\
\hline
\cite{RIS-ID-Rel1}& $\checkmark$ & $\checkmark$ & $\checkmark$ \\
\hline
\end{tabular}
\end{table*}

Consider a wireless communication system with a large number of RISs deployed in a complex urban environment, where a BS aims to serve a specific user equipment (UE) in a specific time slot utilizing available nearby RISs. Here, to the best of the authors’ knowledge, the current RIS literature assumes that the availability of BS-RIS and UEs-RIS links is always guaranteed as a given in the system design. However, due to the mobility of UEs and/or obstacles, these links can appear/disappear instantly. Therefore, the BS needs first to be aware of over which RIS or RISs the UE has the UE-RIS-BS link available before it can start optimizing its phase shifts to serve that UE. Specifically, as is the case in channel estimation, the BS needs to frequently apply an association procedure to pair RISs and their corresponding UEs they are expected to serve, as the first step before using these RISs to assist the BS-UE communication. Following this pairing procedure is expected to lead to more efficient resource allocation and interference management by adaptively reallocating resources (RISs) based on the UEs' needs, channel conditions, and the overall performance metric that needs to be maximized. Investigating this problem gives practical engineering insights into the overall RIS-assisted wireless network operating mechanism by considering the interaction of all its components over time: BS, RISs, UEs, obstacles, and channels.

Against the above background, in this paper,  we consider a system-level problem of a BS aiming to detect and uniquely identify RISs that are reachable by a specific UE in a given time slot. To solve this problem, we propose a novel  RIS identification and detection (RIS-ID) scheme that enables the BS to pair UEs with their corresponding reachable RISs at a specific time slot. Hence, the proposed RIS-ID scheme can be used as an initial and essential step before optimizing each RIS to serve its nearby UE, for more efficient resource allocation. Specifically, the main contribution of this study can be summarized as follows. To the best of the authors' knowledge, this is the first study in the literature to consider the problem of RIS detection and identification by providing a comprehensive mathematical background. To tackle this problem, we propose a novel and simple RIS-ID scheme that does not require hardware modifications at the RIS, UE, or BS side. Furthermore, in order to assess the proposed RIS-ID scheme, we derived the false and miss-detection probabilities as the main performance metrics. Accordingly, we provided a practical procedure to set the detection threshold at the BS side to achieve targeted false and miss-detection probabilities.

Furthermore, building upon this present work, the following few works recently appeared in the literature, which were published earlier than the current work\footnote{This work was online available as a preprint before the other works.}. In \cite{RIS-ID-pract}, the authors conducted a lab experiment to validate the concept of this work in a real-world environment; furthermore, they proposed a novel modulation scheme to enhance the system performance. In \cite{RIS-ID-beam}, the authors extended the proposed RIS-ID scheme to enable the RIS performing both passive beamforming and identification, through a novel RIS elements partitioning algorithm. Finally, in \cite{RIS-ID-Rel1}, the authors extended the proposed RIS-ID scheme by proposing different methods to identify the RIS alternative to the one proposed in this study. In Table I, we summarize the relevance of this work to existing works in the literature.

The rest of the paper is organized as follows. Section II introduces the RIS-ID problem in wireless communication systems, followed by the system model of the proposed scheme. In Section III, we present the fundamental metrics used to evaluate the performance of the proposed scheme: false detection and miss-detection probabilities. Extensive computer simulations and performance analysis are provided in Section IV. Finally, Section V summarizes our findings and discusses potential future directions.

\textit{Notation:} Matrices and column vectors are denoted by an upper and lower case boldface letters, respectively. $\mathbf{X} \in \mathbb{C}^{m \times k}$ denotes a complex-valued matrix $\mathbf{X}$ with $m \times k$ size, where $\mathbf{X}^T$ is the transpose and $[\mathbf{X}]_{n,h}$ is the $(n,h)$-th entry. $\mathbf{0}_N$, $\mathbf{I}_N$, $[ \cdot ]$, $\lceil \cdot \rceil$, and $\mod(\cdot)$ are the $N$-dimensional all-zeros column vector, the $N \times N$ identity matrix, the floor, ceiling, and modulus functions, respectively. $x \sim \mathcal{CN}(0, \sigma^2)$ stands for complex Gaussian distributed random variable (RV) with mean $\mathbb{E}[x] = 0$ and variance $\text{Var}[x] = \sigma^2$.

\section{RIS DETECTION AND IDENTIFICATION} 
In this section, we first introduce the RIS-ID problem considering a general wireless communication system. Afterward, the detailed system model is provided for the proposed RIS-ID solution. 
\subsection{PROBLEM FORMULATION}
\begin{figure*} [t]
    \centering
\includegraphics[width=165mm]{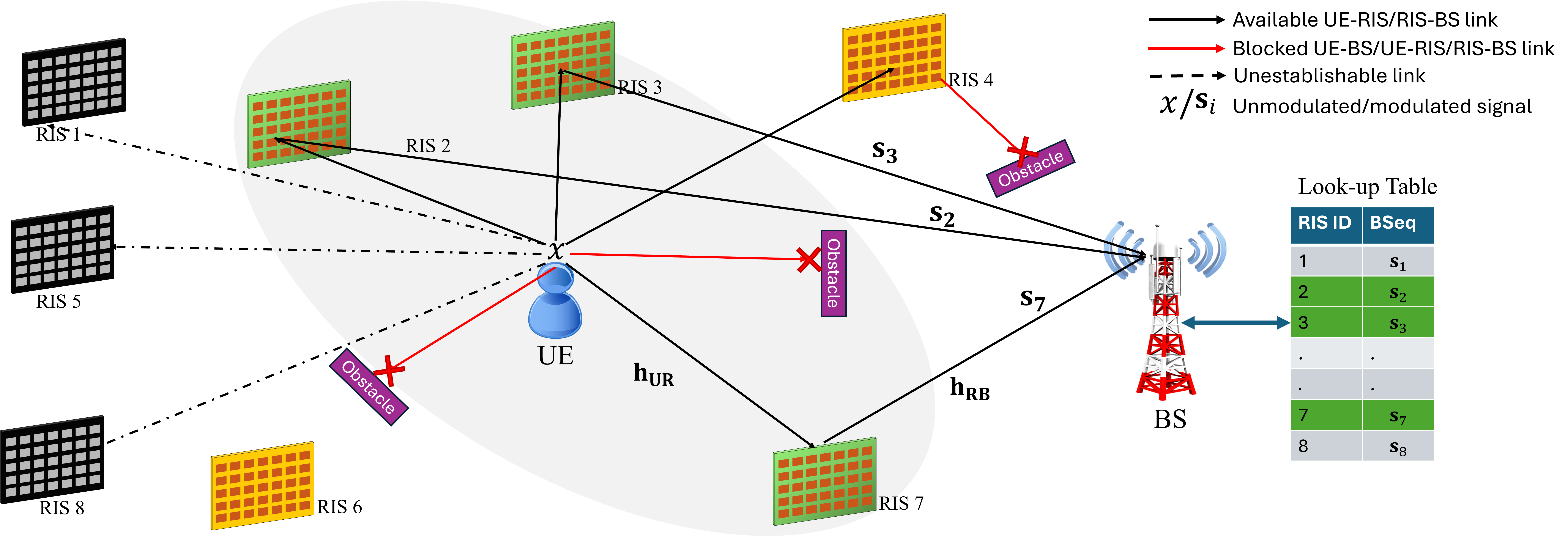}
    \caption{An illustration of an RIS-assisted wireless communication system where a BS aims to detect uniquely identify RISs that are reachable by a specific UE in a specific time slot. There are three reachable (green), two blocked (orange), and three out-of-range RISs (black).}\label{fig:problem-stat}
\end{figure*}

Fig. \ref{fig:problem-stat} shows a wireless communication system with multiple RISs deployed to assist the BS-UE communication. Specifically, in Fig. \ref{fig:problem-stat}, RISs shown in black are the ones out of the range of communication, mainly due to the high path loss associated with the long distance. RISs shown in orange are within the range of communication; however, due to obstacles and UE mobility, the UE-RIS and/or RIS-BS links might be blocked. Finally, RISs shown in green are within the communication range and with the UE-RIS-BS link available. However, due to the mobility of UEs and/or obstacles, these links can appear/disappear instantly. Therefore, the BS needs first to be aware of over which RIS or RISs the UE has the UE-RIS-BS link available before it can start optimizing its phase shifts to serve that UE. In this context, it can be clearly seen that the RIS-ID process is an initial step that comes before channel estimation, as the latter process can be applied only after knowing the availability of the UE-RIS-BS link. Therefore, no channel state information (CSI) is needed for our proposed RIS-ID solution. Here, in a given time slot and for a specific UE, the BS aims to detect and uniquely identify green RISs by utilizing the proposed RIS-ID scheme.  

Following on from the RIS-ID process, the BS can control these green RISs and optimize their phase shifts to assist signal transmission to that specific UE, as conventionally done in the RIS literature. In what follows, a detailed explanation of the proposed RIS-ID scheme is presented.
\subsection{PROPOSED RIS-ID SOLUTION: SYSTEM MODEL}
Consider a single antenna BS\footnote{A single-antenna BS is considered here to simplify the introduction of the proposed RIS-ID problem and its proposed solution, while the extension to multi-antenna BS is straightforward, as the RIS-ID problem and the concept of its proposed solution are the same.} aims to detect and uniquely identify RISs {\it reachable} by a specific UE (green ones in Fig. \ref{fig:problem-stat}) in a given time slot. Each RIS has an ID number that is uniquely associated with its phase shift reflection pattern (PSRP), where a look-up table is used to link RISs' IDs with their corresponding PSRPs. Here, the BS is assumed to know the UE's approximate location within a specific geographical area (gray region in Fig. \ref{fig:problem-stat}) using traditional localization techniques. Accordingly, considering their locations with respect to the gray region, the BS can classify deployed RISs as {\it unreachable} (black ones) and {\it potentially reachable} (red and green ones). 

The RIS-ID process starts when the UE continuously transmits (omnidirectional) an unmodulated carrier signal while the BS receives the signal reflections from reachable RISs modulated with their unique PSRPs\footnote{Here, without loss of generalization for the proposed scheme, we assume all of the RIS elements are continuously being adjusted for the purpose of RIS-ID, or possibly subsurface (partition) of them.}. On the BS side, the RIS-ID process is carried out on the received signal using the look-up table as follows. First, unreachable RISs are excluded from the look-up table due to the high path loss that is expected to prevent establishing the UE-RIS and/or RIS-BS link. Next, the received signal is correlated with locally generated versions of the PSRPs associated with the potentially reachable RISs. Afterward, each PSRP with a correlation output higher than a predetermined detection threshold is mapped back to the corresponding RIS ID, declaring the detection of that RIS as a reachable one. Hence, an optimum output of the RIS-ID process is to declare all (and only) green RISs in Fig. \ref{fig:problem-stat} as reachable RISs.

Let $l$ (superscript) and $m$ (subscript) denote the indices of the RIS and received symbol at the BS side, respectively. Accordingly, let $\mathbf{h}_{\text{UR},m}^{(l)}$ and  $\mathbf{h}_{\text{RB},m}^{(l)} \in \mathbb{C}^{N^{(l)}\times 1}$, denote the UE-RIS  and RIS-BS channels, respectively, where $N^{(l)}$ is the RIS size. Considering spatially correlated Rayleigh fading channels, we have $\mathbf{h}_{\text{UR},m}^{(l)}\sim\mathcal{CN}(\mathbf{0},\beta_{\text{UR}}^{(l)}\mathbf{R})$ and $\mathbf{h}_{\text{RB},m}^{(l)}\sim\mathcal{CN}(\mathbf{0},\beta_{\text{RB}}^{(l)}\mathbf{R}), \forall l,m, $ where $\mathcal{CN}(0, \sigma^2)$ stands for complex Gaussian distribution with zero mean and variance $\sigma^2$. Here, $\beta_{\text{UR}}^{(l)}$ and $\beta_{\text{RB}}^{(l)}$ represent the path gains for the UE-RIS and RIS-BS channels, respectively; $\mathbf{0}$ denotes $N^{(l)}$-dimensional vector of zeros; and $\mathbf{R}\in \mathbb{C}^{N^{(l)}\times N^{(l)}} $ denotes the RIS spatial correlation matrix \cite{bjornson2020rayleigh}. Also, $\mathbf{\Phi}_m^{(l)}\in \mathbb{C}^{N^{(l)}\times N^{(l)}}$ and $n_m\sim\mathcal{CN}(0,\sigma_n^2)$ denote the RIS phase shift matrix and the additive white Gaussian noise (AWGN) sample, respectively. Thus, when there are $L$ potentially reachable RISs (green and orange ones in Fig. \ref{fig:problem-stat}) exist in the vicinity of UE, the received signal corresponding to the $m$-th reflected symbol can be expressed as\footnote{Here, we consider that the propagation delays associated with different RISs are so small that the reflected symbols from all RISs reach the BS in approximately the same time. Note that this is a reasonable assumption due to the geometry of the propagation environment: a close proximity of UE-side RISs employed in the system compared to the long distances between the BS-RISs.}
\vspace{-0.2cm}
\begin{align}
y_m &=x\sum_{l=1}^{L}\sqrt{P}\left[(\mathbf{h}_{\text{UR},m}^{(l)})^T \mathbf{\Phi}_m^{(l)} \mathbf{h}_{\text{RB},m}^{(l)}\right]\eta^{(l)}+n_m \nonumber 
\end{align}
\begin{align}
        &\overset{\text{I}}{=}x\sum_{l=1}^{L}e^{j\phi_{m}^{(l)}}\left[\sqrt{P}(\mathbf{h}_{\text{UR},m}^{(l)})^T  \mathbf{I}_{N^{(l)}} \mathbf{h}_{\text{RB},m}^{(l)}\right] \eta^{(l)}+n_m\nonumber \\
        &\overset{\text{II}}{=}\sum_{l=1}^{L}e^{j\phi_{m}^{(l)}}\tilde{h}^{(l)}\eta^{(l)}+n_m \nonumber\\
         &=\sum_{l=1}^{L}\mathbf{s}_{m}^{(l)}\tilde{h}^{(l)}\eta^{(l)}+n_m, 
        \label{eq:y}
\end{align}
where, $P$ is the transmit power, $x=1$ stands for the complex baseband sample corresponding to the passband transmission of unmodulated carrier signal $\cos(2\pi f_ct)$ with $f_c$ is the operating frequency and $t$ denotes time. Furthermore, in step I, corresponding to the PSRP, we set $\mathbf{\Phi}_m^{(l)}=e^{j\phi_m^{(l)}}\mathbf{I}_{N^{(l)}}$ with unity reflection amplitude, and $\mathbf{I}_{N^{(l)}}$ denotes an identity matrix with size $N^{(l)}$. In step II, considering that the channel coherence time is larger than the time required to receive all of the reflected symbols at BS\footnote{Another option, when the BSeq has a longer duration (fast fading channels), is to correlate only a segment of the received signal such that the channel coherence time is still larger \cite{synch-book}.}, we drop the symbol index $m$ and set $\tilde{h}^{(l)}=\sqrt{P}(\mathbf{h}_{\text{UR}}^{(l)})^T\mathbf{I}_{N^{(l)}} \mathbf{h}_{\text{RB}}^{(l)}$. Furthermore, $\mathbf{s}_m^{(l)}=e^{j\phi_{m}^{(l)}}$ denotes the $m$-th element of the PSRP vector  $\mathbf{s}^{(l)}$ associated with the $l$-th RIS. Here, we use $\eta^{(l)}\in\{0,1\}$ as an RV that indicates if the $l$-th RIS is reachable ($\eta^{(l)}=1$) or unreachable ($\eta^{(l)}=0$), from the UE perspective. Note that, for our proposed RIS-ID scheme, no CSI related to the RIS links is needed.

Here, motivated by the work in \cite{modulated}, the PSRP of the RIS ($\mathbf{\Phi}_m^{(l)}$) is chosen to modulate $x$ with binary phase shift keying (BPSK) symbols of a unique binary sequence (BSeq) vector $\mathbf{q}^{(l)}$, as follows. Independent from the UE, the RIS periodically adjusts its elements with a common phase shift that is applied to all of them,  without the need for synchronization between different RISs. Specifically, at the reflection of the $m$-th symbol, the phase shift applied to $x$ can be expressed as
\begin{align}
\label{phi_m_modified}
\phi_m^{(l)}=\frac{1}{j}\ln (2q_{\tilde{m}}^{(l)}-1),\;m,\tilde{m}\in\{1,2,\dots,M\},
\end{align}
where $\tilde{m}=(c+m-1)\mod M +1$ is the index of the $m$-th BSeq symbol modulating $x$. Here, $c$ is the index of the first BSeq symbol modulating $x$, where it is seen from the BS side as a random integer number; we model it using the uniform distribution, $c\in\mathcal{U}\{1,2,\dots,M\}$. Note that $c$ is modeled as such to reflect the fact that the UE is not in synchronization with RISs, which are independently and periodically adjusting their phases to reflect the BSeq symbols. Therefore, when the UE starts transmitting $x$, the first BSeq symbol modulating $x$ at the RIS side can be any of an $M$ possible  symbols. Furthermore, $q_{\tilde{m}}^{(l)}$ is the $\tilde{m}$-th element (symbol) of the BSeq vector $\mathbf{q}^{(l)}=[q_1^{(l)},\cdots,q_M^{(l)}], q_m^{(l)}\in\{0,1\}$. Note that \eqref{phi_m_modified} corresponds to the phase shift adjustment required to modulate the unmodulated carrier signal sent from the UE according to a specific binary sequence. Therefore, each element's phase shift is adjusted according to (2) so that the unmodulated carrier signal is reflected either with a phase shift of $0$ (representing symbol $1$) or with a phase shift of $\pi$ (representing symbol $-1$).\\
\indent If there is no reachable RIS in the vicinity of the UE, the received signal at BS can be rewritten as 
\begin{align}
y_m=n_m,\label{eq:noisy-y}
\end{align}
after interference from other RF sources is ignored.\\
\indent Overall, the BS collects the reflected BSeq symbols simultaneously from all of the $L$ RISs, where considering the order they are received with, they can be represented by the vector
\begin{align}
\mathbf{y}=[y_1,\dots,y_{v_1},y_{v_1+1},\dots,y_{v_1+M}, \dots,y_{v_1+M+v_2}],\label{eq:y_clctd}  
\end{align}
where, $\mathbf{y}\in \mathbb{C}^{1\times (v_1+M+v_2)}$, $v_1$ and $v_2$ are the numbers of samples collected before and after the received reflections' samples of $x$ from the $L$ RISs, respectively. Here, $v_1$ and $v_2$ are used to show that the BS starts collecting samples earlier than the UE transmission and waits after the end of transmission, to guarantee collecting all of the reflected BSeq symbols. Therefore, these additional collected samples contain the AWGN samples only, where $y_1,\cdots,y_{v_1},y_{v_1+M+1},\cdots,y_{v_1+M+v_2}\sim\mathcal{CN}(0,\sigma^2_n)$, as in \eqref{eq:noisy-y}. Accordingly, the samples $y_{v_1+1},\cdots,y_{v_1+M}$ correspond to the actual received signal samples, $y_m$'s, which were already given in \eqref{eq:y}.

In light of the above, the detection process for the $l$-th RIS starts by generating the detected symbol \cite{synch-book}
\begin{align}
 d_{c,k}^{(l)}=\frac{1}{\sqrt{M}}\sum_{m=1}^{M}\mathbf{s}_{c,m}^{(l)}\mathbf{y}_{m+k},0\leq k\leq v_1+v_2, \label{eq:dck}
\end{align}
 by letting $\mathbf{s}^{(l)}$ denoting the BPSK modulated sequence vector $\mathbf{q}^{(l)}$, we have $\mathbf{s}_{c,m}^{(l)}\in\{-1,1\}$ corresponds to the (${(c+m-1)\; \text{mod}\; M +1}$)-th element of the vector $\mathbf{s}^{(l)}$, or equivalently, the $m$-th element of the vector $\mathbf{s}^{(l)}$ with its elements are circularly shifted by $c$ to the left. Also, $\mathbf{y}_{m+k}$ is the ($m+k$)-th element of $\mathbf{y}$. Note that the normalization factor $1/\sqrt{M}$ is used to reduce the overall noise variance resulting from the correlation process in case of no RIS, leading to significantly high false detection probability. Next, we obtain $c$ and $k$ associated with the maximum correlation amplitude as
\begin{align}
(\hat{c},\hat{k})^{(l)}=\underset{c, k}{\arg\max}\;|d_{c,k}^{(l)}|^2.\label{eq:c_k}
\end{align}
Finally, from \eqref{eq:c_k}, we obtain our decision metric as
\begin{align}
D^{(l)}=|d_{\hat{c},\hat{k}}^{(l)}|^2.\label{eq:D} 
\end{align}
Consequently, an RIS detection is declared only when $D^{(l)}>r^{(l)}$, where $r^{(l)}$ is a predetermined threshold that will be discussed in the next section. Accordingly, from the BS perspective, the $l$-th RIS is considered to be reachable by the UE only if $D^{(l)}>r^{(l)}$, irrespective of the geographical location of that RIS relative to the UE. Note that the estimation of propagation delay (and hence, $v_1$) can be obtained from $\hat{k}^{(l)}$ as it is done in \cite{modulated}, while here, we focus on the RIS detection identification only.
\section{PERFORMANCE METRICS OF THE RIS-ID SCHEME}\label{sec:P_f}
In this section, we provide two fundamental metrics to assess the performance of the proposed RIS-ID scheme. First, we derive the false detection probability $P_\mathrm{F}^{(l)}$, which corresponds to the probability of declaring a detection ($\eta^{(l)}=1$) of the $l$-th RIS while it is unreachable ($\eta^{(l)}=0$). Next, we derive the miss-detection probability $P_\mathrm{miss}^{(l)}$ as the probability of declaring the $l$-th RIS unreachable ($\eta^{(l)}=0$) while it is reachable ($\eta^{(l)}=1$). 
\vspace{-0.35cm}
\subsection{FALSE DETECTION PROBABILITY}
In light of the earlier introduction, the false detection probability for the $l$-th RIS can be written as
\begin{align}
P_\mathrm{F}^{(l)} &= P\!\left(D^{(l)} > r^{(l)} \mid \eta^{(l)} = 0\right) \notag\\
&= P\!\left(D^{(l)} > r^{(l)} \,\middle|\,
\bigcap_{\substack{\tilde l=1 \\ \tilde l \neq l}}^{L-1} \eta^{(\tilde l)}\right)
P\!\left(\bigcap_{\substack{\tilde l=1 \\ \tilde l \neq l}}^{L-1} \eta^{(\tilde l)}\right) \notag\\
&\overset{\text{I}}{=} \sum_{w=1}^{2^{L-1}} B^{(l)} P\!\left(
\bigcap_{\substack{\tilde l=1 \\ \tilde l \neq l}}^{L-1}
(\eta^{(\tilde l)} = \eta^{(\tilde l)}_w)\right) \notag\\
&\overset{\text{II}}{=} \sum_{w=1}^{2^{L-1}} B^{(l)} P\!\left(\eta^{(\tilde l)} = \eta^{(\tilde l)}_w\right) \notag\\
&\overset{\text{III}}{=} \left(\tfrac{1}{2}\right)^{L-1} \sum_{w=1}^{2^{L-1}} B^{(l)} .
\label{eq:P_F-general}
\end{align}
\vspace{-0.6cm}
where,
\begin{equation}
B^{(l)}=P\bigg(D^{(l)}>r^{(l)}\bigg|\bigcap\limits_{\overset{\tilde{l}=1}{\tilde{l}\neq l}}^{L-1}(\eta^{(\tilde{l})}=\eta^{(\tilde{l})}_w)\bigg). 
\vspace{-0.15cm}\label{eq:ff}
\end{equation}
\normalsize{} \noindent In step I of \eqref{eq:P_F-general}, we consider all of the possible combinations of potentially reachable/unreachable RISs given in the look-up table, where we have $w=2^{L-2}\eta^{(L-1)}_w+2^{L-3}\eta^{(L-2)}_w+\dots+\eta^{(1)}_w+1$ possibilities. Without loss of generality, in step II, we consider that the $l$-th RIS can be reachable/unreachable independent from the $\tilde{l}$-th RIS, $\forall l, \tilde{l}\in\{1,\dots,L\}$, and in step III, an RIS is considered to be reachable or unreachable with the same probability\footnote{In practice, for a given RISs deployment environment, these probabilities can be obtained empirically from real-world measurements.}.

In order to proceed, we need to find the right-tail probability associated with $D^{(l)}$, as follows. First, from \eqref{eq:dck}, we expand $d_{c,k}^{(l)}$ as follows
\begin{equation}
\resizebox{\linewidth}{!}{%
$\begin{aligned}
d_{c,k}^{(l)} &= \frac{1}{\sqrt{M}} \sum_{m=1}^{M} \mathbf{s}_{c,m}^{(l)} \mathbf{y}_{m+k}\allowdisplaybreaks \\
&\overset{\text{I}}{=} \frac{1}{\sqrt{M}} \left[ \sum_{m_1=1}^{v_1 -k} \mathbf{s}_{c,m_1}^{(l)} n_{\tilde{m}_1}
+ \hspace{-0.5cm} \sum_{m_2=v_1-k+1}^{M} \hspace{-0.4cm} \mathbf{s}_{c,m_2}^{(l)} \left( \sum_{l=1}^{L} 
\mathbf{s}_{\tilde{m}_2}^{(l)} \tilde{h}^{(l)} \eta^{(l)} + n_{\tilde{m}_2} \right) \right] \\
&\overset{\text{II}}{=} \frac{1}{\sqrt{M}} \left[ \sum_{m=1}^{M} \tilde{n}_m + \tilde{h}^{(l)} A_{c,k}^{(l)} \eta^{(l)} 
+ \sum_{\substack{\tilde{l}=1 \\ \tilde{l} \neq l}}^{L-1} \tilde{h}^{(\tilde{l})} A_{c,k}^{(\tilde{l})} \eta^{(\tilde{l})} \right] \\
&= \frac{1}{\sqrt{M}} \left( A_{c,k}^{(l)} \tilde{h}^{(l)} \eta^{(l)} 
+ \sum_{\substack{\tilde{l}=1 \\ \tilde{l} \neq l}}^{L-1} \tilde{h}^{(\tilde{l})} A_{c,k}^{(\tilde{l})} \eta^{(\tilde{l})} + \tilde{n} \right),
\end{aligned}$%
}
\label{eq:d_c_k_exp}
\end{equation}

\normalsize{}where $\tilde{m}_1=m_1+k, \tilde{m}_2=m_2+k$, $\tilde{n}=\sum_{m=1}^{M}\tilde{n}_m$, and 
\begin{equation}
A_{c,k}^{(d)}=
    \begin{cases}
        \displaystyle \sum_{m_2=v_1-k+1}^{M}\hspace{-0.3cm}\mathbf{s}_{c,m_2}^{(l)}\mathbf{s}_{\tilde{m}_2}^{(d)}, & \text{for } k<v_1,\\
         \displaystyle \sum_{m_2=1}^{M}\hspace{-0.1cm}\mathbf{s}_{c,m_2}^{(l)}\mathbf{s}_{\tilde{m}_2}^{(d)}, & \text{for }  k=v_1,\\
         \displaystyle \sum_{m_2=1}^{M+v_1-k}\hspace{-0.1cm}\mathbf{s}_{c,m_2}^{(l)}\mathbf{s}_{\tilde{m}_2}^{(d)}, &\text{for } k>v_1, 
    \end{cases}
\end{equation}
with $d\in\{l,\tilde{l}\}$. Note that in step I, we isolate pure noise terms from the ones of the received signal and, without loss of generality, we assume $v_1<M$. In step II, we further isolate all of the noise terms from the ones of the pure received signal samples, and the main signal of interest from other interference terms. Consequently, without loss of generality and in light of \eqref{eq:c_k} and \eqref{eq:d_c_k_exp}, we rewrite \eqref{eq:D} as
\begin{align}
    D^{(l)}&=\frac{1}{M}\max\bigg(\bigg|A_{c,k}^{(l)}\tilde{h}^{(l)}\eta^{(l)}+\sum_{\underset{\tilde{l}\neq l}{\tilde{l}=1}}^{L-1}\tilde{h}^{(\tilde{l})}A_{c,k}^{(\tilde{l})}\eta^{(\tilde{l})}+\tilde{n}\bigg|^2\bigg)\nonumber\\
    &\overset{\text{I}}{\approx}\frac{1}{M}\max\bigg(\bigg|A_{c,k}^{(l)}\tilde{h}^{(l)}\eta^{(l)}+\sum_{\underset{\tilde{l}\neq l}{\tilde{l}=1}}^{L-1}\tilde{h}^{(\tilde{l})}A_{c,k}^{(\tilde{l})}\eta^{(\tilde{l})}\bigg|^2\bigg),\label{eq:D-upper}
\end{align}
where, in step I, we ignored the noise effect for large values of $N^{(l)}$ and/or $P\beta^{(l)}$, as the received signal variance is much larger than the one of the noise, as will be shown later in this subsection. As seen from \eqref{eq:D-upper}, it is very difficult to obtain the exact distribution of $ D^{(l)}$\footnote{While not explicitly addressed in this study, the distribution of $ D^{(l)}$ for general $L$ can be obtained through numerical methods.}. Therefore, in what follows, we consider the single ($L=1$) and two ($L=2$) RISs cases to demonstrate the performance metrics of our proposed RIS-ID scheme.
\subsubsection{Single RIS case ($L=1$)}
The probability that the ($l=1$)-th RIS is falsely detected (a single RIS declared to be reachable) can be obtained from \eqref{eq:P_F-general}, as
\begin{align} P_\mathrm{F}^{(1)}=P(D^{(1)}>r^{(1)}),\label{eq:P_F_1} 
\end{align}
where if there is no reachable RIS, we obtain the received signal as given in \eqref{eq:noisy-y}. Note that, for $y_m=n_m$, $\mathbf{s}_{c,m}^{(1)}\in\{-1,1\}$, and considering the random distribution symmetry, we have $d_{c,k}^{(1)}\sim\mathcal{CN}(0,\sigma_n^2)$, as it is a normalized sum of independent and identically distributed (i.i.d.) complex Gaussian random variables (CGRVs). Therefore, we obtain $|d_{c,k}^{(1)}|^2\sim\chi^2_2$, where $\chi^2_z$ denotes the chi-square random distribution with $z$ degrees of freedom. Due to the overlap in the summed elements of different $d_{c,k}^{(1)}$'s, they are dependent RVs, which makes it very challenging to obtain the exact distribution of $D^{(1)}$. Nevertheless, the right-tail probability of $D^{(1)}$ in \eqref{eq:P_F_1} can be upper-bounded as \cite{max-dep}
\begin{equation}
\resizebox{0.8\linewidth}{!}{%
$\begin{aligned}
P_\mathrm{F}^{(1)} &\le \min\left\{1, (v_1 + v_2 + 1) M \left(1 - F(r^{(1)})\right) \right\} \\
&= \min\left\{1, (v_1 + v_2 + 1) M \exp\left(-\frac{r^{(1)}}{\sigma_n^2}\right) \right\},
\end{aligned}$%
}
\label{eq:Pf_upper}
\end{equation}

\noindent\hspace{0cm}where $F(r^{(1)})=1-e^{-r^{(1)}/\sigma_n^2}$ is the cumulative distribution function (CDF) of $|d_{c,k}^{(1)}|^2$ \cite{quadratic-forms} and $(v_1+v_2+1) M$ is the number of arguments (dependent $\chi^2$ RVs) of the $\mathrm{max}$ function in \eqref{eq:c_k}, obtained according to \eqref{eq:dck}.

As seen from \eqref{eq:Pf_upper}, the false detection probability is inversely proportional to the threshold value $r^{(1)}$, while it is directly proportional to the number of collected extra samples ($v_1 +v_2$), the sequence length $M$, and the noise variance $\sigma_n^2$. Accordingly, neither the characteristics of the reflected signal nor the UE-RIS or RIS-BS channels have any effect on the false detection of an RIS.
\subsubsection{Two RISs case ($L=2$)}
From \eqref{eq:P_F-general}, for $L=2$, $P_\mathrm{F}^{(l)}$ can be written as
\begin{align}
\scalebox{0.94}{$
P_\mathrm{F}^{(l)}=\frac{1}{2}\Big(P\big(D^{(l)}>r^{(l)}\mid \eta^{(\tilde{l})}=0\big) 
+ P\big(D^{(l)}>r^{(l)}\mid \eta^{(\tilde{l})}=1\big)\Big)
$}\label{eq:P_F_2}
\end{align}

\noindent \hspace{-0.12cm}where, $P(D^{(l)}>r^{(l)}|\eta^{(\tilde{l})}=0)$ is given in \eqref{eq:Pf_upper}. To derive the second probability, we note that for $\eta^{(\tilde{l})}=1, \eta^{(l)}=0$ and using \eqref{eq:D-upper}, we obtain
\vspace{-0.15cm}
\begin{align}
    D^{(l)}&\approx\frac{1}{M}\max\big(\big|A_{c,k}^{(\tilde{l})}\tilde{h}^{(\tilde{l})}\big|^2\big)=\frac{|\tilde{h}^{(\tilde{l})}|^2}{M}\max\big(\big|A_{c,k}^{(\tilde{l})}\big|^2\big)\nonumber\\
    &=H^{(\tilde l)}\big|A_{\hat{c},\hat{k}}^{(\tilde{l})}\big|^2,
    \label{eq:D_Pfalse}
\end{align}
where $H^{(\tilde l)}=\frac{|\tilde{h}^{(\tilde{l})}|^2}{M}$.

To obtain the distribution of $H^{(\tilde l)}$, first, we obtain the distribution of the sum $\tilde{h}^{(\tilde{l})}=\sum_{i=1}^{N^{(\tilde{l})}}\sqrt{P}(h_{\text{UR},i}^{(\tilde{l})} h_{\text{RB},i}^{(\tilde{l})})$, where $h_{\text{UR},i}^{(\tilde{l})}$ and $h_{\text{RB},i}^{(\tilde{l})}$ are $i$-th components of channels $\mathbf{h}_{\text{UR}}^{(\tilde{l})}\sim\mathcal{CN}(0,\beta_{\text{UR}}^{(\tilde{l})}\mathbf{R})$ and $\mathbf{h}_{\text{RB}}^{(\tilde{l})}\sim\mathcal{CN}(0,\beta_{\text{RB}}^{(\tilde{l})}\mathbf{R})$, respectively. Note that the product of the two complex Gaussian RVs, $h_{\text{UR},i}^{(\tilde{l})}$ and $ h_{\text{RB},i}^{(\tilde{l})}$, results in a complicated distribution. In addition, due to the spatial correlation between RIS elements, we have $(h_{\text{UR},i}^{(\tilde{l})} h_{\text{RB},i}^{(\tilde{l})})$'s are non-independent RVs, making it challenging to obtain the exact distribution of $\tilde{h}^{(\tilde{l})}$. Therefore, to make the derivations more tractable, we first ignore the spatial correlations between RIS elements, $\mathbf{R}=\mathbf{I}_{N^{(\tilde{l})}}$; and thus, $(h_{\text{UR},i}^{(\tilde{l})} h_{\text{RB},i}^{(\tilde{l})})$'s are i.i.d. RVs\footnote{We show later in simulation results that the effect of spatial correlation is insignificant, validating our i.i.d. approximation.}. In light of this, we use the central limit theorem (CLT) to approximate the distribution of the overall sum $\tilde{h}^{(\tilde{l})}=\sum_{i=1}^{N^{(\tilde{l})}}\sqrt{P}(h_{\text{UR},i}^{(\tilde{l})} h_{\text{RB},i}^{(\tilde{l})})$ to the CG distribution, irrespective of the distribution of its individual elements. 
We have $\mathbb{E}[\sqrt{P}(h_{\text{UR},i}^{(\tilde{l})} h_{\text{RB},i}^{(\tilde{l})})]=0$ and $\text{Var}[\sqrt{P}(h_{\text{UR},i}^{(\tilde{l})} h_{\text{RB},i}^{(\tilde{l})})]=P\beta^{(\tilde{l})}$, where $\beta^{(\tilde{l})}=\beta_{\text{UR}}^{(\tilde{l})}\beta_{\text{RB}}^{(\tilde{l})}$. Accordingly, we obtain $\tilde{h}^{(\tilde{l})}\sim\mathcal{CN}(0,N^{(\tilde{l})}P\beta^{(\tilde{l})})$ as $N^{(\tilde{l})}\rightarrow \infty$. Therefore, $H^{(\tilde l)}=|\tilde{h}^{(\tilde{l})}|^2/M$ is a chi-square RV with two degrees of freedom ($\chi^2_2$). Furthermore, due to the random cyclic shift ($c$), $\big|A_{\hat{c},\hat{k}}^{(\tilde{l})}\big|^2$ becomes a discrete RV with finite number of distinct values. Consequently, $D^{(l)}$ corresponds to the product of a continuous and discrete RVs and its right-tail probability can be found as   
\begin{equation}
\resizebox{\linewidth}{!}{%
$\begin{aligned}
&P(D^{(l)}>r^{(l)}|\eta^{(\tilde{l})}=1) = 1 - P(D^{(l)}<r^{(l)}|\eta^{(\tilde{l})}=1) \\
&\approx 1 - P(H^{(\tilde l)} |A^{(\tilde l)}_{\hat{c},\hat{k}}|^2 < r) \\
&= 1 - \sum_{w=1}^W P(H^{(\tilde l)}a_w^{(\tilde l)} < r \big| |A^{(\tilde l)}_{\hat{c},\hat{k}}|^2 = a_w^{(\tilde l)}) 
P(|A^{(\tilde l)}_{\hat{c},\hat{k}}|^2 = a_w^{(\tilde l)}) \\
&\overset{\text{I}}{=} 1 - \sum_{w=1}^W \left(1 - \exp\left(\frac{-r^{(l)}}{\frac{N^{(\tilde l)}P\beta^{(\tilde l)} a^{(\tilde l)}_w}{M}}\right) \right) 
P(|A^{(\tilde l)}_{\hat{c},\hat{k}}|^2 = a^{(\tilde l)}_w),
\end{aligned}$%
}
\label{eq:P_F_2_2}
\end{equation}

\noindent\hspace{0cm}where, $0\leq a^{(\tilde l)}_w\leq M^2$ is an integer. In step I, we substituted the CDF of the chi-square RV $H^{(\tilde l)}a_w^{(l)}$ \cite{simon2002probability}. Finally, the probability given in \eqref{eq:P_F_2} can be written in terms of \eqref{eq:Pf_upper} and \eqref{eq:P_F_2_2} as
\vspace{-0.2cm}
\begin{align}
    P_\mathrm{F}^{(l)}&\approx \frac{1}{2}\min{\{1,(v_1+v_2+1) M\exp(-r^{(l)}/\sigma_n^2})\}\nonumber\\
    &+\frac{1}{2}-\frac{1}{2}\sum_{w=1}^W \left(1-\exp\left(\frac{-r^{(l)}}{\frac{N^{(\tilde l)}P\beta^{(\tilde l)} a^{(\tilde l)}_w}{M}}\right) \right)\nonumber\\ &\times P(|A^{(\tilde 
 l)}_{\hat{c},\hat{k}}|^2=a^{(\tilde l)}_w)\nonumber\\
 &\overset{\text{I}}{\approx} \frac{1}{2}-\frac{1}{2}\sum_{w=1}^W \left(1-\exp\left(\frac{-r^{(l)}}{\frac{N^{(\tilde l)}P\beta^{(\tilde l)} a^{(\tilde l)}_w}{M}}\right) \right)\nonumber\\ &\times P(|A^{(\tilde 
 l)}_{\hat{c},\hat{k}}|^2=a^{(\tilde l)}_w),\label{eq:P_F_2_3}
\end{align}
where, in step I, we considered that the second probability (associated with the event where the $\tilde l$-th RIS is reachable, $\eta^{(\tilde l)}=1$) is dominating the first one associated with the noise samples only ($\eta^{(\tilde l)}=0$), due to large 
 $P\beta^{(\tilde l)}$ and/or $N^{(\tilde l)}$ values.
\subsection{MISS-DETECTION PROBABILITY}
By following the same derivation steps given in Section \ref{sec:P_f}-A, the miss-detection probability can be obtained, by using \eqref{eq:P_F-general} and considering the right-tail probability (switching the inequality sign), as
\begin{align}
P_\mathrm{miss}^{(l)}=&P(D^{(l)}<r^{(l)}|\eta^{(l)}=1)\nonumber\\
    \overset{\text{III}}{=}&\left(\frac{1}{2}\right)^{L-1}\sum_{w=1}^{2^{L-1}}P\bigg(D^{(l)}<r^{(l)}\bigg|\bigcap\limits_{\overset{\tilde{l}=1}{\tilde{l}\neq l}}^{L-1}(\eta^{(\tilde{l})}=\eta^{(\tilde{l})}_w)\bigg),\label{eq:P_miss-general}
\end{align}
where, $w$ is given in Section \ref{sec:P_f}-A.

In what follows, we consider the miss-detection probability of the two cases: $L=1$ and $L=2$.
\subsubsection{Single RIS case ($L=1$)} Considering the $(l=1)$-th RIS as the only reachable RIS and using \eqref{eq:P_miss-general}, the probability of miss-detecting it (declaring it unreachable) can be given as
\begin{align}
    P_\mathrm{miss}^{(1)}=P(D^{(1)}<r^{(1)}).\label{eq:P_miss_1}
\end{align}
Note that, when $\eta^{(1)}=1$ and using \eqref{eq:D-upper}, we have
\begin{align}
 D^{(1)}&\approx\frac{1}{M}\max\bigg(\bigg|A_{c,k}^{(1)}\tilde{h}^{(1)}\bigg|^2\bigg)=\frac{|\tilde{h}^{(1)}|^2}{M}\max\bigg(\bigg|A_{c,k}^{(1)}\bigg|^2\bigg)\nonumber\\
&=\frac{|\tilde{h}^{(1)}|^2|A_{\hat{c},\hat{k}}^{(1)}|^2}{M} \overset{\text{I}}{=}|\sqrt{M}\tilde{h}^{(1)}|^2,
\end{align}

\noindent where $D^{(1)}\sim\chi^2(2)$ as shown previously. In step I, we considered the fact that for the optimal parameters $\hat{c}$ and $\hat{k}$, we have perfect alignment between $\mathbf{s}_{c,m_2}^{(1)}$ and $\mathbf{s}_{\tilde{m}_2}^{(1)}$ with $A_{\hat{c},\hat{k}}^{(1)}=M$ corresponding to the maximum correlation amplitude. Consequently, $P_\mathrm{miss}^{(1)}$ corresponds to the CDF of $D^{(1)}$ and can be given as
\begin{align}
    P_\mathrm{miss}^{(1)}\approx1-\exp\left(\frac{-r^{(1)}}{N^{(1)}P\beta^{(1)}M}\right).
    \label{eq:P_miss_1}
\end{align}
As in the $P_\mathrm{F}^{(1)}$, $P_\mathrm{miss}^{(1)}$ is directly affected by the threshold setting; however, it is directly proportional to $r^{(l)}$ and inversely to $M, N^{(1)}$ $P$ and $\beta^{(1)}$.
\subsubsection{Two RISs case ($L=2$)}
Considering the $l$-th RIS be reachable ($\eta^{(l)}=1$) and using \eqref{eq:P_miss-general}, we have
\begin{align}
    \scalebox{0.93}{
    $P_\mathrm{miss}^{(l)}=\frac{1}{2}\Big(P\big(D^{(l)}<r^{(l)}|\eta^{(\tilde{l})}=0\big)+P\big(D^{(l)}<r^{(l)}|\eta^{(\tilde{l})}=1\big)\Big)$},\label{eq:P_miss_2}
\end{align}

\normalsize{} \noindent where the first probability is given in \eqref{eq:P_miss_1}. To obtain the second probability, we note that, for $\eta^{(l)}=\eta^{(\tilde {l})}=1$ and using \eqref{eq:D-upper}, we obtain
\begin{align}
 D^{(l)}&\approx\frac{1}{M}\max\big(\big|A_{c,k}^{(l)}\tilde{h}^{(l)}+A_{c,k}^{( \tilde l)}\tilde{h}^{(\tilde{l})}\big|^2\big)\nonumber\\
&\overset{\text{I}}\le \frac{1}{M} \max\big(\big(\big|A_{c,k}^{(l)}\tilde{h}^{(l)}\big|+\big|A_{c,k}^{( \tilde l)}\tilde{h}^{(\tilde{l})}\big|\big)^2\big)\nonumber\\ 
&\overset{\text{II}}\le \frac{1}{M}\big(A_{\hat{c},\hat{k}}^{(l)}\big|\tilde{h}^{(l)}\big| + \tilde A^{(\tilde l)}\big|\tilde{h}^{(\tilde{l})}\big|\big)^2\nonumber\\
&\overset{\text{III}}{=} \frac{1}{M}\big(M\big|\tilde{h}^{(l)}\big| + \tilde A^{(\tilde l)}\big|\tilde{h}^{(\tilde{l})}\big|\big)^2,\label{eq:D_p_miss_2} 
\end{align}
where, in step I, we used $|a+b|^2=|a|^2+|b|^2+2\mathbb{R}\{ab^*\}\leq|a|^2+|b|^2+2|a||b|=(|a|+|b|)^2$. Furthermore, in step II, we considered the upper bounds for the two terms, $A_{c,k}^{(l)}\leq A_{\hat{c},\hat{k}}$ and $A_{c,k}^{(\tilde{l})}\leq \tilde{A}^{(\tilde{l})},\forall c,k$, where we obtain $\tilde{A}^{(\tilde{l})}$ empirically through simulations, while, in step III, we set $A_{\hat{c},\hat{k}}=M$ denoting the maximum possible correlation amplitude. Consequently, using \eqref{eq:D_p_miss_2}, we obtain
\begin{equation}
\resizebox{\linewidth}{!}{%
$\begin{aligned}
&P(D^{(l)}\le r^{(l)}|\eta^{(\tilde{l})}=1)\ge P\bigg(\frac{1}{M}\big(M|\tilde{h}^{(l)}| + \tilde A^{(\tilde{l})}|\tilde{h}^{(\tilde{l})}|\big)^2\le r^{(l)}\bigg)\\
&= P\left(\left|\frac{M\tilde{h}^{(l)}}{\sqrt{M}}\right| + \left| \frac{\tilde A^{(l)} \tilde{h}^{(\tilde{l})}}{\sqrt{M}} \right| \le \sqrt{r^{(l)}}\right) \overset{\text{I}}{=}P(R\le \sqrt{r^{(l)}})\\
&\overset{\text{II}}{=}\frac{1}{2}-\int_{0}^{\infty} \frac{\Im\{e^{jw\sqrt{r^{(l)}}}\Psi_{R}(w)\}}{w\pi}dw,
\end{aligned}$%
}
\label{eq:P_miss_2_1}
\end{equation}

\normalsize{} \noindent where in step I, we have $R=R_1+R_2$ with $R_1=\left|\frac{M\tilde{h}^{(l)}}{\sqrt{M}}\right|$ and $R_2=\left| \frac{\tilde A^{(\bar l)}\tilde{h}^{(\tilde{l})}}{\sqrt{M}} \right|$ are Rayleigh distributed RVs. In step II, we utilize Gil-Pelaez’s
inversion formula to obtain the CDF of $R$ using its characteristic function (CF)\cite{mathai1992quadratic}, which is given as
\begin{align}
    \Psi_{R}(w)=\Psi_{R_1}(w)\Psi_{R_2}(w),
\end{align}
where $\Psi_{R_1}(w)$ and $\Psi_{R_2}(w)$ are the CFs of $R_1$ and $R_2$, respectively, and for $i={1,2}$, we have 
\begin{align}
    \Psi_{R_i}(w)=e^{-\frac{w^2\sigma_i^2}{2}} \sum_{k=0}^{\infty} \frac{\left(\frac{1}{2}w^2\sigma_i^2\right)^k}{(2k-1)!} + j\sqrt{\frac{\pi}{2}}w\sigma_ie^{-\frac{w^2\sigma_i^2}{2}},
\end{align}
with $\sigma_1=\sqrt{MN^{(l)}P\beta^{(l)}}$ and $\sigma_2=\sqrt{\frac{\tilde A^{(\bar l)} }{M}N^{(\tilde l)}P\beta^{(\tilde l)}}$. Finally, substituting \eqref{eq:P_miss_1} and \eqref{eq:P_miss_2_1} in \eqref{eq:P_miss_2}, we get
\begin{equation}
\resizebox{\linewidth}{!}{%
$\begin{aligned}
P_\mathrm{miss}^{(l)} \geq \frac{3}{2}
- \exp\left(\frac{-r^{(l)}}{N^{(l)} P \beta^{(l)} M} \right)
- \int_{0}^{\infty} \frac{\Im\{ e^{j w \sqrt{r^{(l)}}} \Psi_{R}(w) \}}{w \pi} \, dw.
\end{aligned}$%
}
\label{eq:P_miss_2RIS}
\end{equation}
\subsection{DETECTION THRESHOLD SETTING}
Here, in light of the previous mathematical analyses, we discuss the threshold setting at the BS side to achieve a predetermined performance level.  
\subsubsection{Single RIS case ($L=1$)}
For the BS to set the detection threshold, first, the required false detection probability $P_\mathrm{F}^{(l)}$ value needs to be set as a given design parameter. Furthermore, the noise variance $\sigma^2_n$ needs to be obtained using actual signal measurements at the automatic gain control (AGC) output or as a constant value from its data sheet \cite{synch-book}. Using the obtained $\sigma^2_n$ and the required $P_\mathrm{F}^{(l)}$, the detection threshold can be obtained from \eqref{eq:Pf_upper}. It is worth noting that, as can be seen from \eqref{eq:Pf_upper} and \eqref{eq:P_miss_1}, the detection threshold $r^{(l)}$ can not be set to both minimize $P_\mathrm{F}^{(l)}$ and $P_\mathrm{miss}^{(l)}$. Furthermore, $r^{(l)}$  and sequence length $M$ are  the only parameters to control $P_\mathrm{F}^{(l)}$, while $N^{(l)}$ and $P$  can also be used in the case of $P_\mathrm{miss}^{(l)}$, as follows.

Having the threshold and sequence length set and in order to obtain the required miss-detection probability, we  solve \eqref{eq:P_miss_1} for $N^{(l)}$ to obtain
\begin{align}
\bar{N}^{(l)}=\frac{-r^{(l)}}{P\beta^{(l)}M\ln(1-P_\mathrm{miss}^{(l)})},
\end{align}
where $\bar{N}^{(l)}$ corresponds to the required RIS size to obtain the targeted $P_\mathrm{miss}^{(l)}$. Likewise, \eqref{eq:P_miss_1} can be solved for $P$ to obtain the required transmit power for the targeted $P_\mathrm{miss}^{(l)}$.

\subsubsection{Two RISs case ($L=2$)} 
In the two RISs case, as seen from \eqref{eq:P_F_2_3} and \eqref{eq:P_miss_2RIS}, all parameters affect both $P_\mathrm{F}^{(l)}$ and $P_\mathrm{miss}^{(l)}$. Specifically, $P_\mathrm{F}^{(l)}$ is directly proportional to $P$ and $N^{(l)}$, and inversely to $r^{(l)}$ and $M$. In contrast, $P_\mathrm{miss}^{(l)}$ is directly proportional to $r^{(l)}$ and inversely to $P$, $N^{(l)}$, and $M$. Furthermore, it can be noted that $P_\mathrm{F}^{(l)}$ and $P_\mathrm{miss}^{(l)}$ exhibit an opposite behavior in response to the change in $P$, $N^{(l)}$, and $r^{(l)}$, while they have a similar inverse proportion to $M$. Consequently, in order to set the detection threshold, the two functions ($P_\mathrm{F}^{(l)}$ and $P_\mathrm{miss}^{(l)}$) can be plotted (y-axis) against a range of threshold values (x-axis) to select the value that satisfies both $P_\mathrm{F}^{(l)}$ and $P_\mathrm{miss}^{(l)}$ requirements, as will be shown in Section \ref{Sim_results}.

\section{SIMULATION RESULTS} \label{Sim_results}
In this section, we provide extensive computer simulations to assess the performance of the proposed RIS-ID scheme, focusing on evaluating two critical performance metrics: false detection and miss-detection probabilities. Without loss of generality, as a BSeq, Walsh-Hadamard codes are used to represent RISs IDs in the look-up table. This is because these codes are known for their widespread use in code division multiple access (CDMA) applications owing to their orthogonality property, which aligns with our purpose of distinguishing multiple RISs simultaneously. Furthermore, we assume that all RISs have identical sizes, as well as identical UE-RIS and RIS-BS distances (and thus path gains). Therefore, the index $l$ is omitted from $N$, $\beta$, and $r$. Note that, without loss of generality, assuming identical distances for both RISs corresponds to a maximum mutual interference scenario, representing a conservative approach to show the performance of the proposed RIS-ID scheme. Otherwise, considering different distances, the BS can apply successive interference cancellation to detect the two BSeqs sequentially, mitigating the mutual interference effectively. We also consider an operating bandwidth ($\text{BW}$) of $20$ MHz, the number of extra collected samples $v_1+v_2=0.25M$ where $ M $ is the sequence length and $v_1\in\mathcal{U}\{1,2,\dots,0.25M\}$, and $\sigma^2_{n}=-174$ dBm $+ 10\log_{10} (\text{BW})=-101$ dBm \cite{liu2020rach_noise}. 

The overall UE-RIS-BS path gain, $\beta$, is calculated as \cite{Path_Loss}
\begin{align}
\beta=\frac{\lambda^4}{256\pi^2(d_{\text{UR}})^2(d_{\text{RB}})^2},
\end{align}
where $\lambda$ represents the wavelength corresponding to $f_c = 1.8$ GHz; $d_{\text{UR}} = 10$ m and $d_{\text{RB}} = 50 $m are the RIS-UE and RIS-BS distances, respectively. Also, we consider the channel model defined in Section II-B.

The spatial correlation matrix $\mathbf{R}$ is constructed based on the model proposed in \cite{bjornson2020rayleigh}. Accordingly, the spatial correlation between the $w$-th and $\tilde w$-th RIS elements is expressed as 
\begin{align}
    [\mathbf{R}]_{w, \tilde{w}}=\operatorname{sinc}\left(\frac{2\left\|\mathbf{u}_w-\mathbf{u}_{\tilde{w}}\right\|}{\lambda}\right),  \;w,\tilde{w}\in\{1,2,\dots,N\},
\end{align}
where $\operatorname{sinc}(\tau)=\sin (\pi \tau) /(\pi \tau)$ is the sinc function, $\mathbf{u}_k= \left[0, \quad i_\mathrm{H}(w)d_\mathrm{H}, i_\mathrm{V}(w) d_\mathrm{V}\right]^T$ is the spatial vector representing the position of the $w$-th RIS element, with $i_\mathrm{H}(w)=\bmod \left(w-1, N_\mathrm{H}\right)$, $i_\mathrm{V}(w)=\left\lfloor(w-1) / N_\mathrm{H}\right\rfloor$, $d_\mathrm{H}$ and $d_\mathrm{V}$ correspond the horizontal index, vertical index, width, and length of the $w$-th element, respectively. Here, $N_\mathrm{H}$ denotes the number of RIS elements per row. In what follows, we provide simulation results for the three cases ($L=1$, $L=2$ and $L=5$), separately.

\begin{figure}
    \centering
    \includegraphics[width=70mm]{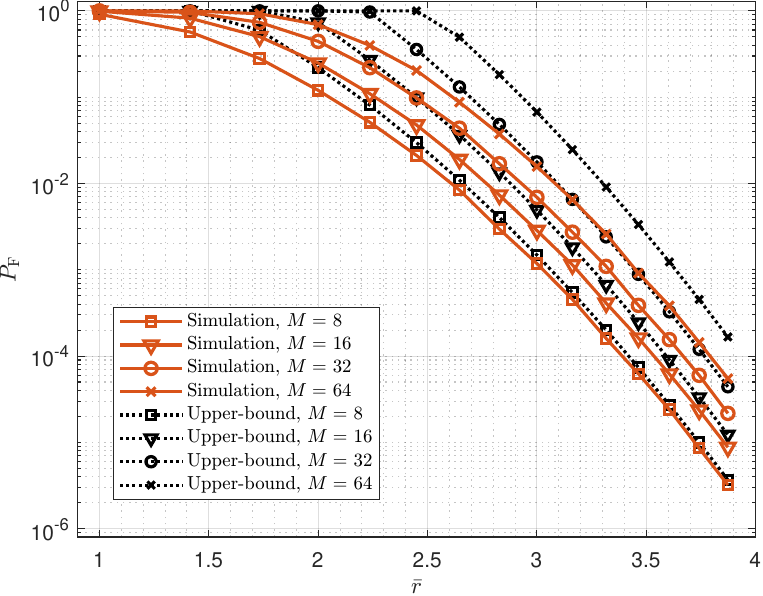}
    \caption{The impact of the sequence length $M$  on the false detection probability when $L=1$.}\label{fig:PF}
\end{figure}
\subsection{SINGLE RIS CASE}
In this subsection, we consider a single potentially reachable RIS by an UE, where the BS aims to detect its presence and uniquely identify it. Here, we set $N=64$ and $M=16$, unless otherwise stated. 

In Fig. \ref{fig:PF}, we investigate the influence of the sequence length $M$ on the false detection probability $P_\mathrm{F}$, where the x-axis indicates the normalized threshold value, $\bar r=(r/\sigma_n^2)^{1/2}$.
In \eqref{eq:Pf_upper}, $(v_1+v_2+1) M$ is the number of dependent RVs of the $\mathrm{max}$ function in \eqref{eq:c_k}. However, some of these RV's are identical since we use a Walsh-Hadamard code whose circularly shifted versions, for different shift values, are the same or negatives of each other. To obtain the refined theoretical upper bound shown in Fig. \ref{fig:PF}, we replaced the number of dependent RVs, $(v_1+v_2+1)M$, with the number of distinct dependent RVs, given by $(v_1+v_2+1)M\rho$, where $\rho = 0.5$ for the Walsh-Hadamard code used in the simulation. It can be seen that obtained theoretical upper bound is tight, where it converges to the simulation curve at higher values of $\bar r$ and lower values of $M$. It can be seen that obtained theoretical upper bound is tight, where it converges to the simulation curve at higher values of $\bar r$ and lower values of $M$. Furthermore, increasing $M$ increases $P_\mathrm{F}$, as more noise samples are involved in the correlation process and, thus, more arguments in \eqref{eq:c_k} to select from. For instance, when $\bar r$ is set to $3$, doubling $M$ leads to a $2.4$-fold increase in the $P_\mathrm{F}$.

\begin{figure}[t!]
    \centering
    \includegraphics[width=70mm]{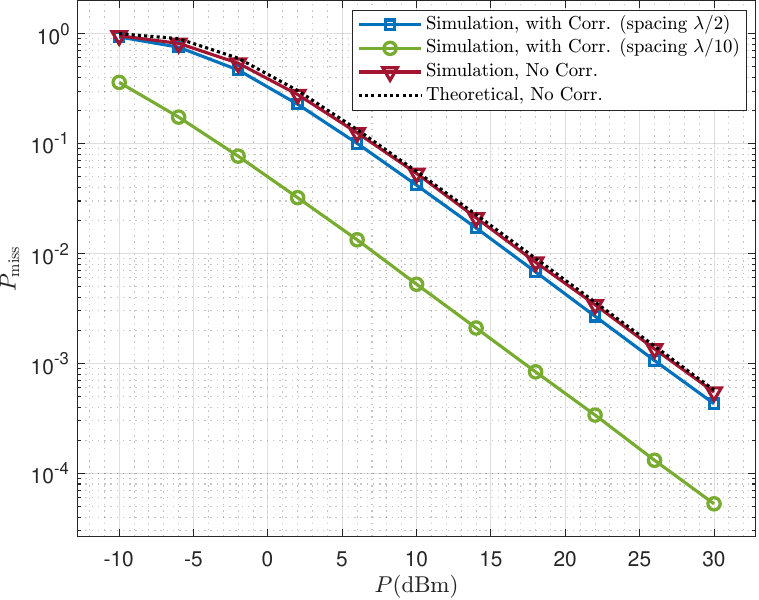}
    \caption{Miss-detection probability with and without correlation, for $L=1$.}\label{fig:Pmiss_withCorr}
\end{figure}
Fig. \ref{fig:Pmiss_withCorr} illustrates  the $P_\mathrm{miss}$ for the proposed scheme in the presence and absence of the spatial correlation between RIS elements. Here, $\bar r$  is set to $3$, which upper-bounds $P_\mathrm{F}$ to $0.005$ as shown in Fig. \ref{fig:PF}. It can be seen that the spatial correlation among RIS elements ($\lambda/2$) does not significantly impact the performance of the proposed scheme where $P_\mathrm{miss}$ is only around $1.25$ times higher when considering it. Furthermore, it can be noted that the simulation curve for the uncorrelated scenario aligns closely with the theoretical one obtained under the no-correlation assumption, demonstrating the validity of our derivations in Section \ref{sec:P_f}. Finally,  in the high correlation case ($\lambda/10$), the $P_\mathrm{miss}$ performance is shown to be enhanced significantly. This is due to the high correlation amplitude resulting from the superposition of the highly correlated reflected paths that are inherently more aligned in phase.                         
\begin{figure}[t!]
    \centering
    \includegraphics[width=70mm]{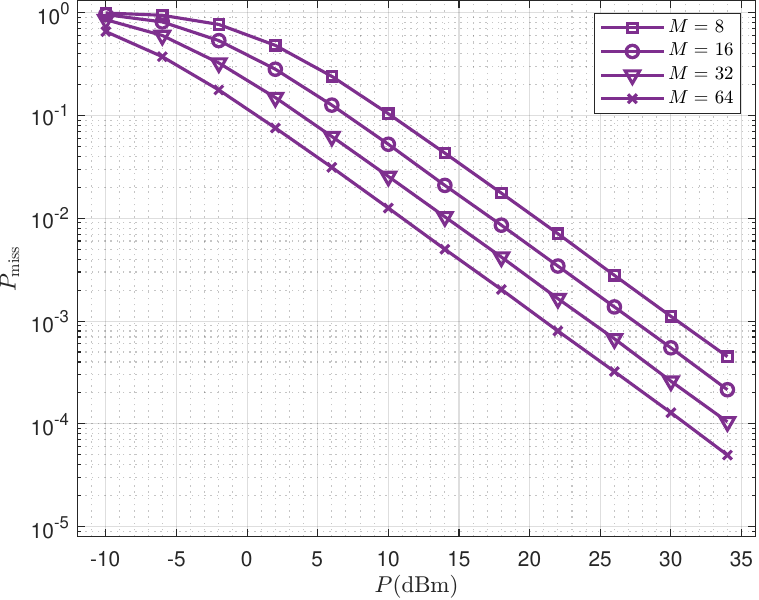}
    \caption{The impact of $M$ on the miss-detection probability when $L=1$.}\label{fig:Pmiss_DifferentM}
\end{figure}

In Fig. \ref{fig:Pmiss_DifferentM}, we show the impact of the sequence length $M$  on the miss-detection probability $P_\mathrm{miss}$ for $\bar r=3$. It can be seen that increasing $M$ decreases the $P_\mathrm{miss}$, where doubling $M$ yields approximately $3$ dB gain for the required $P$ to achieve $P_\mathrm{miss}=10^{-2}$. This behavior can be verified from \eqref{eq:P_miss_1}, where $P_\mathrm{miss}$ inversely proportional to $M$. It is worth noting that while the increase in $M$ leads to an improvement for the $P_\mathrm{miss}$, it concurrently results in performance degradation for the $P_\mathrm{F}$, as seen in Fig. \ref{fig:PF}. Therefore, when determining the optimal value of $M$, it is imperative to consider the required values for $P_\mathrm{F}$ and $P_\mathrm{miss}$ together, as discussed in Section \ref{sec:P_f}-C. 

\begin{figure}[t!]
    \centering
    \includegraphics[width=70mm]{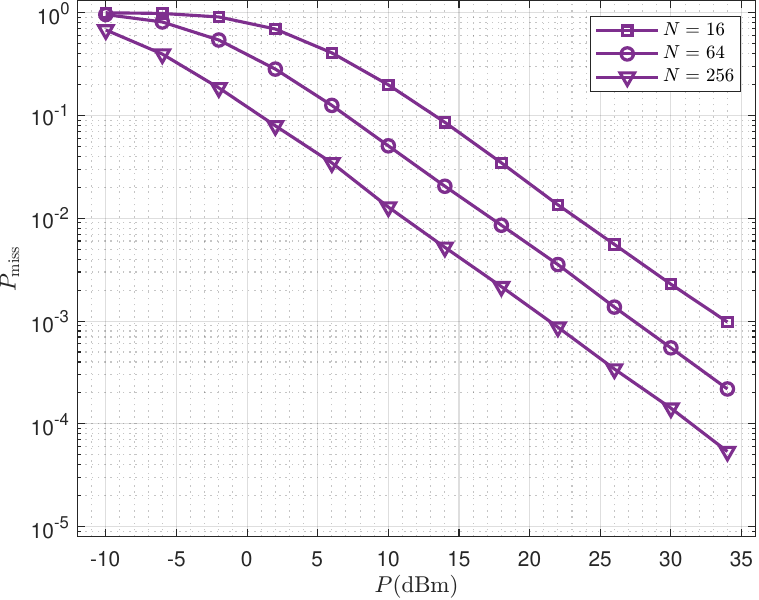}
    \caption{The impact of the RIS size on the miss-detection probability when $L=1$.}\label{fig:Pmiss_DifferentN}
\end{figure}

In Fig. \ref{fig:Pmiss_DifferentN}, we examine the $P_\mathrm{miss}$ for various RIS sizes when $\bar r=3$. It can be seen that as $N$ increases, there is an improvement in the $P_\mathrm{miss}$. For instance, at a $P_\mathrm{miss}$ value of $10^{-2}$, around $6$ dB gain is achieved for the required $P$ by quadrupling $N$. This behavior can be verified from \eqref{eq:P_miss_1}, where $P_\mathrm{miss}$ is inversely proportional to $N$.
\subsection{TWO RISs CASE} \label{Sim_results_TwoRIS}
In this case, we consider two potentially reachable RISs by UE, denoted as RIS 1 and RIS 2, where the BS aims to identify reachable ones. Throughout this subsection, unless explicitly stated, we have $M=32$ and $N=256$, while $P_\mathrm{F}$ and $P_\mathrm{miss}$ probabilities are obtained for the first RIS ($l=1$). In what follows, we first investigate the effects of certain parameters (such as $M$ and $N$) on $P_\mathrm{F}$ and $P_\mathrm{miss}$. Subsequently, we present an illustration chart of the system's overall performance, encompassing performance metrics for both RISs.

\begin{figure}[t!]
    \centering
    \includegraphics[width=70mm]{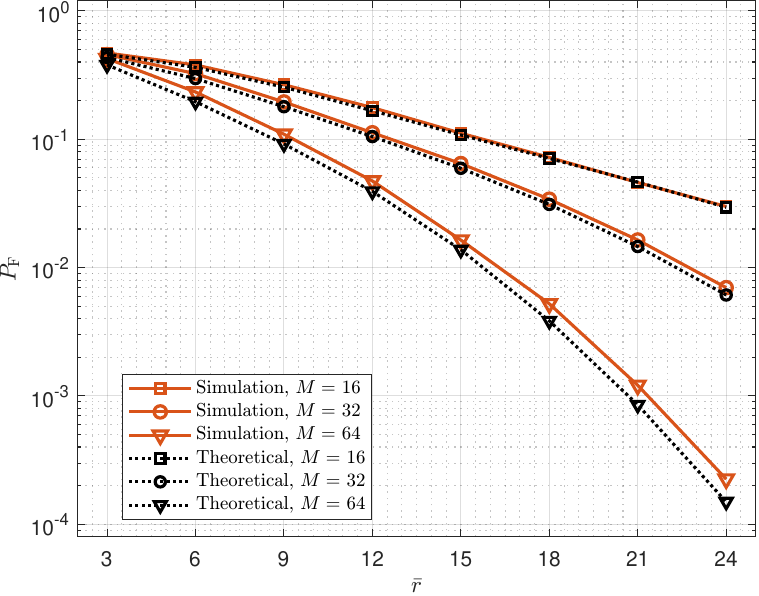}
    \caption{The impact of $M$ on the false detection probability when $L=2$. }\label{fig:TwoRISs_PFalse_forRIS1_forM}
\end{figure}

In Fig. \ref{fig:TwoRISs_PFalse_forRIS1_forM}, we show the effect of the sequence length $M$ on $P_\mathrm{F}$ when $P$ is set to $25$ dBm. First, we observe that computer simulation curves closely match with the theoretical ones obtained using \eqref{eq:P_F_2_3}. It is noteworthy that the alignment between simulation and theoretical curves is more evident when employing lower values of $M$. This is expected as the variance of the noise term, $M\sigma_n^2$, decreases proportionally with $M$, making our approximation in ignoring the noise term in deriving \eqref{eq:D-upper} more accurate. It can also be seen that, contrary to the $L=1$ case, increasing $M$ decreases $P_\mathrm{F}$. Specifically, when $\bar r$ is set to $18$, increasing $M$ from $16$ to $32$ reduces $P_\mathrm{F}$ by a factor of $2$. This behavior can be verified from \eqref{eq:P_F_2_3} where $P_\mathrm{F}$ is inversely proportional to $M$. Moreover, it is valuable to note that $P_\mathrm{F}$ observed in Fig. \ref{fig:TwoRISs_PFalse_forRIS1_forM} is much higher than the one in Fig \ref{fig:PF}. For instance, when  $M=32$ and $\bar r=3$,  $P_\mathrm{F}$ in Fig. \ref{fig:TwoRISs_PFalse_forRIS1_forM} is around $0.5$ while $P_\mathrm{F}$ in Fig. \ref{fig:PF} is around $0.007$.  This difference validates the accuracy of our assumption to ignore the first probability term in \eqref{eq:P_F_2} (corresponds to $P_\mathrm{F}$ in Fig. \ref{fig:PF}) to approximate $P_\mathrm{F}$ (in Fig. \ref{fig:TwoRISs_PFalse_forRIS1_forM}) as in  \eqref{eq:P_F_2_3}.

\begin{figure}[t!]
    \centering
    \includegraphics[width=70mm]{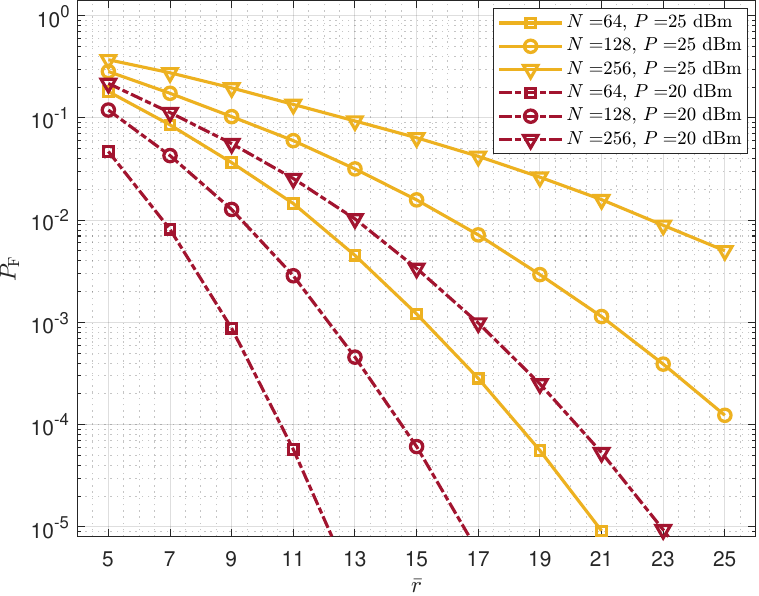}
    \caption{ The false detection probability for $L=2$, varying with different $N$ and $P$ values.}\label{fig:TwoRISs_PFalse_forRIS1_forN}
\end{figure}

In Fig. \ref{fig:TwoRISs_PFalse_forRIS1_forN}, $P_\mathrm{F}$ is evaluated when the RIS size $N$ and UE transmit power $P$ change. It is clearly seen that higher values of $N$ lead to an increase in $P_\mathrm{F}$. As an example, with $P$ is set to $25$ dBm and $\bar r$ is set to $15$, $P_\mathrm{F}$ takes on the values of $0.001$, $0.016$, and $0.064$ for $N=64$, $N=128$, and $N=256$, respectively. Furthermore, it is evident from Fig. \ref{fig:TwoRISs_PFalse_forRIS1_forN} that increasing $P$
results in an increase in $P_\mathrm{F}$, where for $N=256$ and $\bar r=15$, $P_\mathrm{F}$ takes on the values $0.003$ and $0.064$, for $P=20$ dBm and  $P=25$ dBm, respectively. It should be noted that the performance degradation of $P_\mathrm{F}$ with the increment in $P$ and/or $N$ arises from the fact that, as $P$ and/or $N$ increases, the power of the received signal reflected from the $\tilde l$-th RIS increases, increasing the correlation amplitude in \eqref{eq:D}. This behavior can also be verified from \eqref{eq:P_F_2_3}, where $P_\mathrm{F}$ is proportional to $P$ and $N$.       

\begin{figure}
    \centering
    \includegraphics[width=70mm]{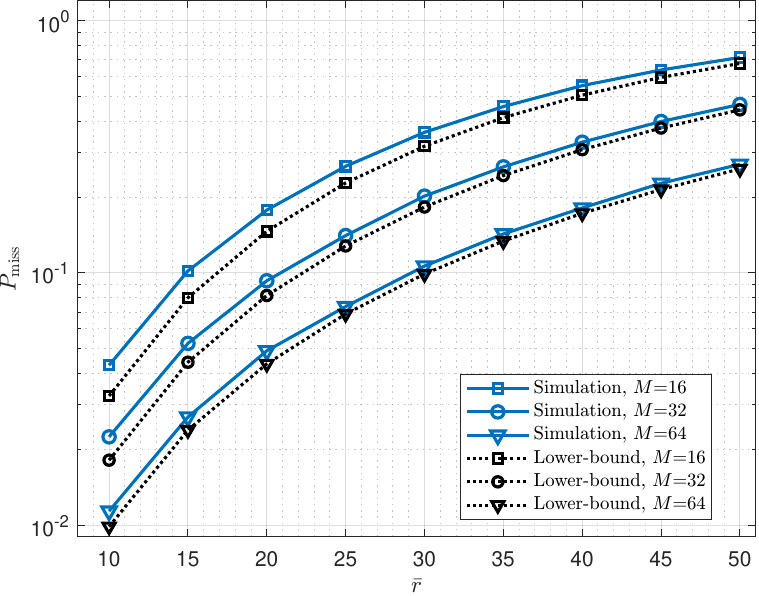}
    \caption{The impact of $M$ on the miss-detection probability when $L=2$.}\label{fig:TwoRISs_PMiss_SimvsTheo_forM}
\end{figure}

Fig. \ref{fig:TwoRISs_PMiss_SimvsTheo_forM} shows $P_\mathrm{miss}$ versus normalized threshold $\bar r$, considering different sequence lengths $M$, with $P$ is set to $15$ dBm. First, it can be seen that the theoretical lower-bound derived in \eqref{eq:P_miss_2RIS} closely aligns with the simulation results, where increasing $M$ and $\bar{r}$ values makes this alignment more pronounced. Also, we note that increasing $M$ yields improvement in the $P_\mathrm{miss}$ where doubling $M$ corresponds to approximately a two-fold reduction in $P_\mathrm{miss}$. This note is consistent with the inverse proportional relationship of $P_\mathrm{miss}$ and $M$ expressed in \eqref{eq:P_miss_2RIS}. Moreover, increasing $\bar r$ decreases $P_\mathrm{miss}$, for instance, when $M=32$, $P_\mathrm{miss}$ is $0.053$, $0.141$, $0.263$ for $\bar r=15$, $\bar r=25$, and $\bar r=35$, respectively. This behavior can be verified from \eqref{eq:P_miss_2RIS}, where $P_\mathrm{miss}$ is proportional to $r$.   
\begin{figure}
    \centering
    \includegraphics[width=70mm]{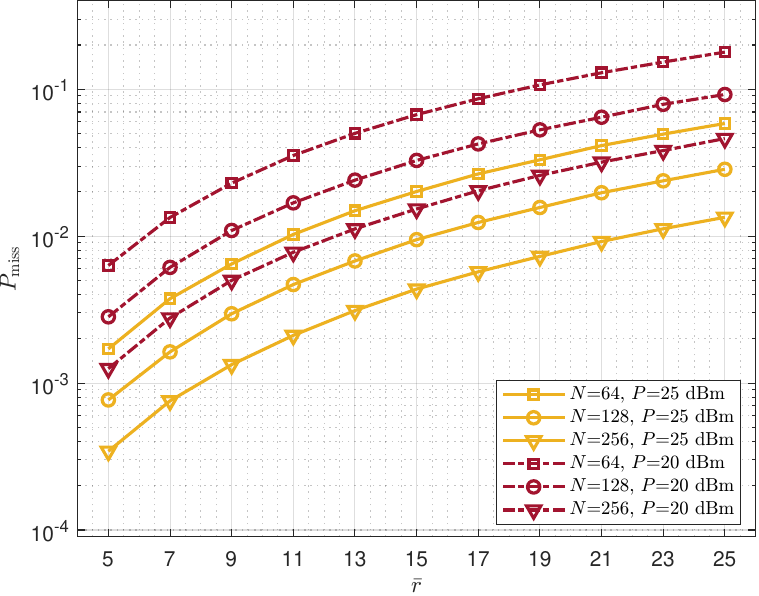}
    \caption{ The miss-detection probability for $L=2$, varying with different $N$ and $P$ values.}\label{fig:TwoRISs_PMiss_forN}
\end{figure}

In Fig. \ref{fig:TwoRISs_PMiss_forN}, we examine the impact of the RIS size $N$ and UE transmit power $P$ on $P_\mathrm{miss}$. It can be observed that, as in the single RIS case, increasing $N$ and/or $P$ both decreases $P_\mathrm{miss}$. For example, with $P=25$ dBm and $\bar r=15$, $P_\mathrm{miss}$ is halved by doubling $N$. Likewise, with $N=256$ and $\bar r=15$, increasing $P$ from $20$ dBm to $25$ dBm decreases $P_\mathrm{miss}$ from $0.015$ to $0.004$. Our observations from Fig. \ref{fig:TwoRISs_PMiss_forN} can be verified from \eqref{eq:P_miss_2RIS}, where $P_\mathrm{miss}$ is inversely proportional to $N$ and $P$.  

\begin{figure}[t!]
    \centering
    \includegraphics[width=70mm]{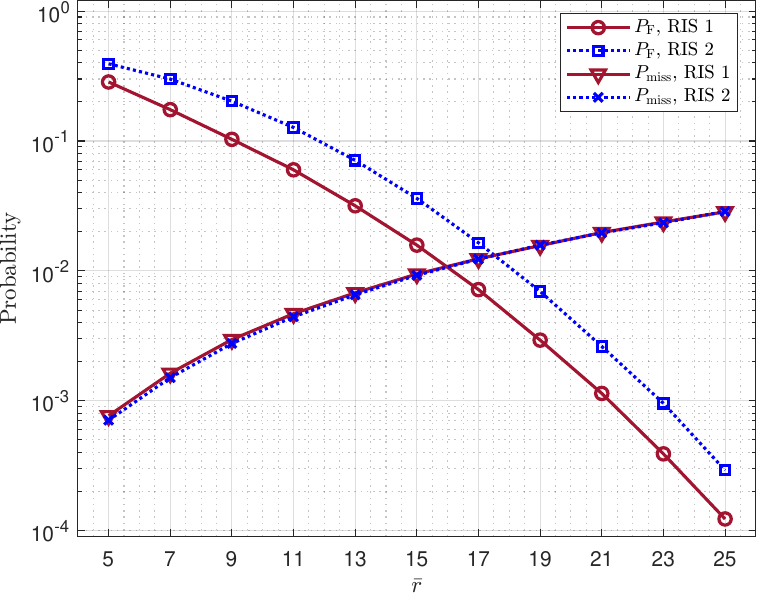}
    \caption{ The impact of threshold setting on $P_\mathrm{F}$ and $P_\mathrm{miss}$, for $P=10$ dBm, $M=32$, $N=128$, and $L=2$.}\label{TWORIS_Pmiss_Pfalse_RIS0RIS1_Together.pdf}
\end{figure}

In Fig. \ref{TWORIS_Pmiss_Pfalse_RIS0RIS1_Together.pdf}, we explore the performance trade-off in terms of the probabilities  $P_\mathrm{miss}$ and  $P_\mathrm{F}$. Differing from previous figures within this section, we incorporate $P_\mathrm{miss}$ and $P_{\mathrm{F}}$ for both RIS 1 and RIS 2. Initially, we see that as $\bar r$ increases, $P_\mathrm{F}$ decreases while $P_\mathrm{miss}$ increases. Moreover, as depicted in Fig. \ref{TWORIS_Pmiss_Pfalse_RIS0RIS1_Together.pdf}, different $P_\mathrm{F}$ and $P_\mathrm{miss}$ values are noticed for different RISs, although they share the same simulation parameters. This discrepancy arises from the different realizations of the discrete RV $|A^{(\tilde l)}_{\hat{c},\hat{k}}|$ in \eqref{eq:P_F_2_3} for different RISs. Furthermore, we observe that the gap between $P_\mathrm{miss}$ curves is much narrower than that in $P_\mathrm{F}$ curves. The reason is that the dominating term in \eqref{eq:P_F_2_3} is the part including $|A^{(\tilde l)}_{\hat{c},\hat{k}}|$, while the dominating term in \eqref{eq:P_miss_2RIS}, common for both RISs, does not involve $A^{(\tilde l)}_{\hat{c},\hat{k}}$.

\begin{figure}[t!]
\centering

\subfloat[$\bar{r}=13$.]{%
  \label{fig:ConfusionCharts_a}%
  \includegraphics[width=43mm]{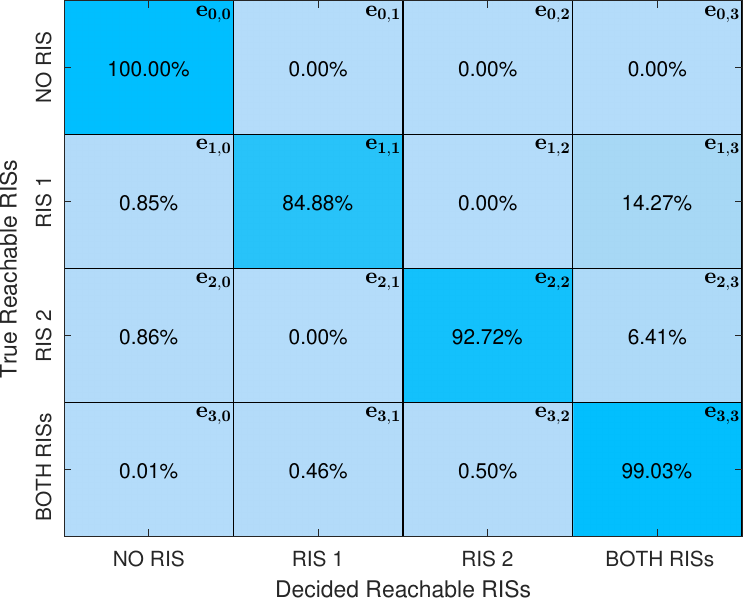}%
}\\

\subfloat[$\bar{r}=17$.]{%
  \label{fig:ConfusionCharts_b}%
  \includegraphics[width=42mm]{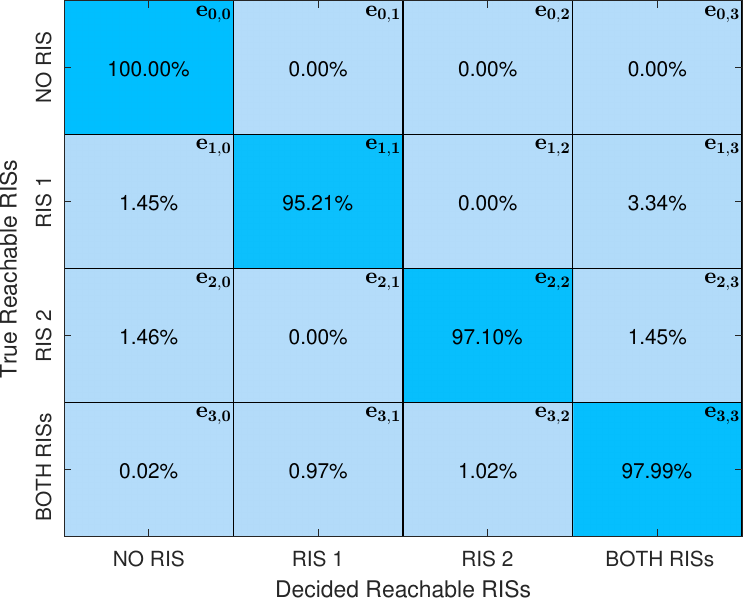}%
}%
\hfill
\subfloat[$\bar{r}=21$.]{%
  \label{fig:ConfusionCharts_c}%
  \includegraphics[width=42mm]{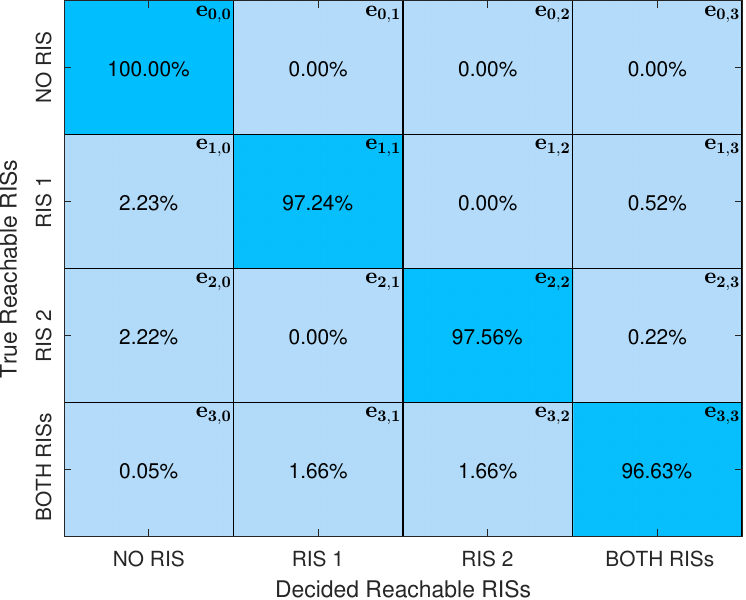}%
}

\caption{Performance evaluation of the proposed RIS-ID scheme using a confusion matrix illustrating the reachability states of RISs for $L=2$, $P=25$ dBm, $M=32$ and $N=128$. The rows and columns represent the true and decided reachability states, respectively, showcasing potential outcomes: ``NO RIS'' ($\eta^{(1)},\eta^{(2)}=0$), ``RIS 1'' ($\eta^{(1)}=1,\eta^{(2)}=0$), ``RIS 2'' ($\eta^{(1)}=0,\eta^{(2)}=1$), and ``BOTH RISs'' ($\eta^{(1)},\eta^{(2)}=1$).}

\label{fig:ConfusionCharts}
\end{figure}

Fig. \ref{fig:ConfusionCharts} summarizes the results in the previous figures in a different illustration approach, using confusion matrix charts for comprehensive comparisons. Here, we provide simulations of $10^7$ random realizations to generate these charts. In each realization, the reachability states of RISs are randomly set with equal probabilities, $P(\eta^{(l)}=0)=P(\eta^{(l)}=1), \forall l$. Then, the proposed RIS-ID scheme is used to decide which RISs are reachable by the UE and which are not. Entries of the confusion matrix are denoted by $e{_i,_j}$ for $i,j=0,\dots, 3$, arranged from top to bottom and left to right, as illustrated in Fig. \ref{fig:ConfusionCharts}. The diagonal entries of the confusion matrix indicate correct decision probabilities regarding RISs' reachability. The off-diagonal entries can be used to find miss-detection and false-detection probabilities for RISs: $P_\text{miss}^{(1)}=[(e_{1,0}+ e_{1,2})/2 + (e_{3,0}+e_{3,2})/2]$, $P_\mathrm{miss}^{(2)}=[(e_{2,0}+ e_{2,1})/2 + (e_{3,0}+e_{3,1})/2]$,  $P_\mathrm{F}^{(1)}=[(e_{0,1}+ e_{0,3})/2 + (e_{2,1}+e_{2,3})/2]$, and $P_\mathrm{F}^{(2)}=[(e_{0,2}+ e_{0,3})/2 + (e_{1,2}+e_{1,3})/2]$.
\subsection{MULTIPLE RISs CASE} \label{Sim_results_MultiRIS}

\begin{figure}[t!]
    \centering
    \includegraphics[width=70mm]{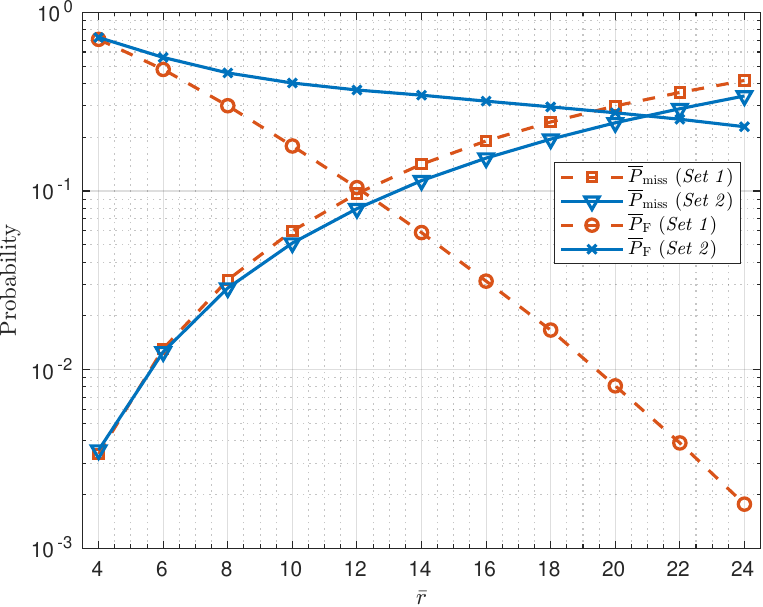}
    \caption{The average miss-detection and false detection probabilities for two different sets of Walsh-Hadamard codes used in the generation of RISs' PSRPs, when $L=5$, $M=16$, $P=15$~dBm, and $N=128$.}\label{fig:5RISs}
\end{figure}

In this subsection, we consider the case where five RISs are potentially reachable by the UE, and the BS aims to identify the reachable ones. To quantify performance, we define the average miss-detection probability as $\overline{P}_{\mathrm{miss}} = \tfrac{1}{L}\sum_{l=1}^{L} P_{\mathrm{miss}}^{(l)}$ and the average false detection probability as $\overline{P}_\mathrm{F} = \tfrac{1}{L}\sum_{l=1}^{L} P_\mathrm{F}^{(l)}$. Here $P_{\mathrm{miss}}^{(l)}$ and $P_\mathrm{F}^{(l)}$ denotes the miss-detection and false detection probabilities of the $l$-th RIS, respectively. 

In Fig. \ref{fig:5RISs}, we evaluate $\overline{P}_{\mathrm{miss}}$ and $\overline{P}_\mathrm{F}$ for two different sets of Walsh-Hadamard codes used in the generation of PSRPs of RISs. As observed, when \emph{Set~1} is used, both $\overline{P}_{\mathrm{miss}}$ and $\overline{P}_\mathrm{F}$ are around $0.1$ at $\bar{r}=12$. However, when \emph{Set~2} is used, we can not simultaneously achieve low values for both probabilities at any $\bar{r}$ value. This performance difference is due to the fact that the codes in \emph{Set~1} have lower cross-correlation than those in \emph{Set~2}, particularly in the presence of time and circular shifts. Therefore, the choice of Walsh–Hadamard codes used to generate the PSRPs of RISs plays an important role, as it considerably affects the overall system performance. Furthermore, Fig. \ref{fig:5RISs} shows that increasing the normalized threshold $\bar{r}$ leads to an increase in $\overline{P}_{\mathrm{miss}}$ while decreasing $\overline{P}_\mathrm{F}$. These behaviors are consistent with the analytical expressions in \eqref{eq:P_F-general} and \eqref{eq:P_miss-general}, where $P_\mathrm{F}$ decreases and $P_{\mathrm{miss}}$ increases as $\bar{r}$ increases.

\section{CONCLUSION}
In this study, 
considering dynamic link blockages, we aim to enable the BS to perceive the UE-RIS potential associations, for better resource allocation. In this context, we introduced the RIS-ID problem and proposed a novel RIS-ID solution to enable the BS to detect and uniquely identify RISs reachable by a UE as an initial and essential step to utilize them effectively for serving this UE. Furthermore, false and miss-detection probabilities have been proposed as metrics to examine the performance of the proposed solution. The theoretical and simulation results clearly show the effectiveness of the proposed RIS-ID scheme when considering different system settings. 

Specifically, our key findings can be summarized as follows. The proposed RIS-ID scheme can be used for any $L$ number of RISs. The performance dominant factor affecting the false and miss-detection probabilities is the length $M$ and orthogonality of the used BSeqs. Without loss of generality, for the considered two RISs case, simulation results show that the proposed RIS-ID scheme can correctly identify the reachability status of the two RISs with a minimum accuracy of $95\%$ and a maximum error percentage of $3\%$ to miss identifying it.   

A potential challenge for the RIS-ID is the orthogonality of the used binary sequences when a large number of RISs
is considered, as it directly affects the correlation amplitude, and thus, the detection probability at the BS side. As a
solution, future research can investigate the use of a time division multiple access scheme in combination with BSeqs to enhance the orthogonality in the time domain and limit the interference between different RISs.

\bibliographystyle{IEEEtran}
\bibliography{Bibliography}
\end{document}